%% file: main.tex
\documentclass[a4paper, conference]{IEEEtran}

\usepackage[T1]{fontenc}

\IEEEoverridecommandlockouts

\usepackage{amsmath,amssymb,amsfonts}
\usepackage{algorithmic}
\usepackage{graphicx}
\usepackage{textcomp}
\usepackage{xcolor}
\usepackage[capitalise]{cleveref}
\usepackage{siunitx}
\usepackage{subfig}
\usepackage{xcolor}
\usepackage{array}
\usepackage{multicol}
\usepackage{multirow}
\usepackage{tabularx}
\usepackage{tabularray}
\usepackage{colortbl}

\definecolor{blue}{rgb}{0.0, 0, 0}
\usepackage[backend=biber, sorting=none, bibstyle=ieee,citestyle=numeric-comp]{biblatex}
\small{
    \bibliography{references.bib}
}    
\DefineBibliographyStrings{english}{%
  mathesis = {MSc. Thesis},
}

\begin{document}

\title{Design and Evaluation of a Multi-Agent Perception System for Autonomous Flying Networks}

\author{
	\IEEEauthorblockN{
	Diogo Ferreira, Pedro Ribeiro, André Coelho, Rui Campos}
	\IEEEauthorblockA{INESC TEC and Faculdade de Engenharia, Universidade do Porto, Portugal\\
    up202006808@edu.fe.up.pt, \{pedro.m.ribeiro, andre.f.coelho, rui.l.campos\}@inesctec.pt}
}

\maketitle

\begin{abstract}
Autonomous Flying Networks (FNs) are emerging as a key enabler of on-demand connectivity in dynamic and infrastructure-limited environments. However, current approaches mainly focus on UAV placement, routing, and resource management, neglecting the autonomous perception of users and their service demands---a critical capability for zero-touch network operation. 

This paper presents the Multi-Agent Perception System (MAPS), a modular and scalable system that leverages multi-modal large language models (MM-LLMs) and agentic Artificial Intelligence (AI) to interpret visual and audio data collected by UAVs and generate Service Level Specifications (SLSs) describing user count, spatial distribution, and traffic demand. MAPS is evaluated using a synthetic multimodal emergency dataset, achieving user detection accuracies above 70\% and SLS generation under 130 seconds in 90\% of cases. Results demonstrate that combining audio and visual modalities enhances user detection and show that MAPS provides the perception layer required for autonomous, zero-touch FNs.
\end{abstract}


\input{Chap_1/Introduction}

\input{Chap_2/chap2}

\input{Chap_3/chap3}

\input{Chap_4/chap4}

\input{Chap_last/Conclusion}

\printbibliography

\end{document}

%% file: Chap_1/Introduction.tex
\section{Introduction\label{sec:Introduction}}

\begin{figure}[!ht]
  \begin{center}
    \includegraphics[width=1.0\columnwidth]{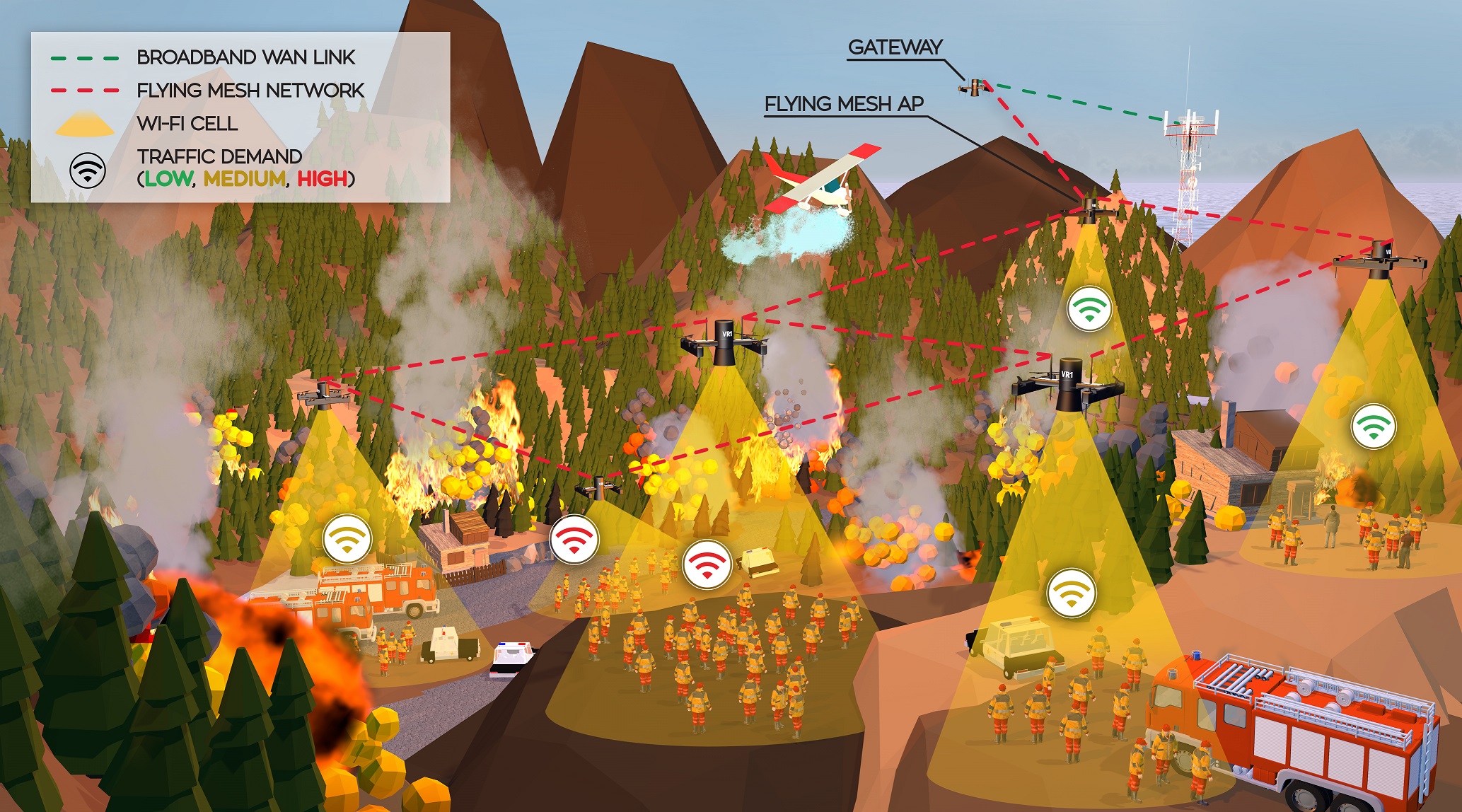}
    \caption{Illustrative example of a Flying Network providing on-demand wireless connectivity to first responders in a disaster management scenario~\cite{Fire_situation}. The UAVs act as aerial access points, extending coverage and enabling communication in the absence of fixed infrastructure.}
    \label{fig:example_scenario}
  \end{center}
\end{figure}

The demand for high-throughput, low-latency communications in dynamic and infrastructure-limited environments has driven research into network architectures capable of rapid deployment and autonomous operation. Flying Networks (FNs), composed of Unmanned Aerial Vehicles (UAVs) equipped with communications payloads, have emerged as a promising solution, providing on-demand scalability and line-of-sight connectivity in scenarios such as disaster response and large public events, as illustrated in Fig.~\ref{fig:example_scenario}. Existing research in FNs primarily addresses network configuration, including UAV placement, routing, and energy optimization, achieving progress in efficiency and service continuity. However, these approaches rely on static assumptions or operator intervention and lack autonomous perception—for example, most FN optimization models assume prior knowledge of user locations, which is unrealistic in highly dynamic environments such as disaster management. This limitation prevents FNs from evolving into fully autonomous systems.

Emerging standards such as the ETSI Zero-touch network and Service Management (ZSM) reference architecture~\cite{ETSI-ZSM002} and the ITU Autonomous Networks framework~\cite{ITU-Y3061} define closed-loop automation and intent-driven orchestration for 6G. However, they assume that high-level service requirements are available, without specifying how these should be derived from real-world perception.

In~\cite{A4FN}, we proposed A4FN, an Agentic Artificial Intelligence (AI) architecture that leverages Large Language Models (LLMs) to enable intent-driven automation in FNs. A4FN consists of two conceptual AI-driven components: a Perception Agent, responsible for semantically interpreting multimodal inputs from UAV-mounted sensors to infer real-time Service Level Specifications (SLSs), and a Decision-and-Action Agent, which uses the inferred SLSs to autonomously reconfigure the FN through UAV repositioning and network reconfiguration. While A4FN highlights the potential of multimodal perception and LLM-based reasoning for autonomous FN control, it remains a conceptual framework; it does not implement the perception pipeline and an approach to derive SLSs through environmental awareness.

To address these gaps, this paper proposes the Multi-Agent Perception System (MAPS), a modular and scalable perception framework designed to provide autonomous environmental awareness in FNs. MAPS leverages Multi-Modal Large Language Models (MM-LLMs) and Agentic AI to interpret visual and audio data collected by UAVs, estimating user count, spatial distribution, and traffic demand. The resulting information is structured into SLSs that feed downstream decision-making agents responsible for UAV control, resource allocation, and network configuration. To the best of our knowledge, MAPS is the first system to fuse multimodal large language models and agentic AI for communications-oriented UAV perception, thereby operationalizing the sensing, data-collection and analytics functions proposed in the autonomous networks and zero-touch management frameworks envisioned by ITU and ETSI.

The main contributions of this paper are three-fold:
\begin{itemize}
    \item \textbf{Multi-agent perception system for FNs:} MAPS autonomously extracts user-centric data (spatial distribution, traffic demand, and contextual cues) from multimodal inputs, providing the perception capability required to enable zero-touch control and management in FNs.
    \item \textbf{Synthetic multimodal dataset:} A dataset integrating visual and audio modalities, generated through a reproducible workflow that can be adapted to a wide range of scenarios. It is intended as a reusable resource for advancing research on multimodal perception in UAV-assisted communication systems, particularly in scenarios where real-world data scarcity is a limiting factor.
    \item \textbf{Performance evaluation:} Validation of MAPS using the synthetic multimodal dataset, confirming its accuracy and responsiveness for near real-time operation in autonomous Flying Networks.
\end{itemize}

This paper is organized as follows. Section~\ref{chap:Related_works} reviews related work on FNs, multimodal perception, and Agentic AI. Section~\ref{chap:MAPS} presents the concept, design, and implementation of MAPS. Section IV describes the synthetic multimodal dataset used for evaluation. Section V presents and analyzes the experimental results. Section VI discusses key findings, limitations, and potential improvements. Finally, Section VII outlines the main conclusions and directions for future work.

%% file: Chap_2/chap2.tex
\section{Related Work}\label{chap:Related_works}

While UAVs offer flexibility and broad coverage, they face challenges such as limited payload capacity and flight autonomy. Early research efforts focused on optimizing UAV placement to meet Quality of Service (QoS) requirements, using algorithms such as SLICER and SUPPLY~\cite{Slicer,1_ribeiro_supply_2024}, which compute QoS-aware positions and energy-efficient trajectories. However, these approaches rely on static inputs and prior knowledge of user distributions, which restricts their autonomy and adaptability to dynamic environments.

To evolve from operator-driven control toward fully autonomous FNs, perception capabilities must be integrated into the UAV network architecture. AI has emerged as a key enabler to reduce or eliminate operator intervention. Specifically, recent advances in Large Language Models (LLMs) have established a new paradigm for autonomous reasoning and perception in complex systems.
Multi-Modal Large Language Models (MM-LLMs) play a particularly relevant role due to their ability to analyze and reason across multiple data modalities, including audio, textual, and visual inputs. Benchmarks spanning structured data interpretation, multidisciplinary reasoning, and real-world perception have been widely adopted to evaluate these models. Zhang et al.~\cite{zhang2024mmllmsrecentadvancesmultimodal} provide a comprehensive review of MM-LLMs, highlighting their expanding capabilities in document understanding, diagram interpretation, and analysis of real-world scenarios.

Despite these advances, MM-LLMs require structured pre-processing and significant computational resources. They are not inherently action-capable and lack integration with the control and orchestration layers essential for communications systems. Consequently, combining MM-LLMs with modular agent-based architectures has emerged as a promising direction in autonomous AI. These modular approaches integrate perception, reasoning, and action in a coordinated manner. Xi et al.~\cite{AI_agents} formalized this paradigm, and subsequent works demonstrated its potential by integrating LLMs with object detection and planning for UAV patrol agents~\cite{Patrol_Agent}. Still, while promising, such systems do not address communication-oriented perception or user-centric demand estimation.

Recent studies have explored the integration of LLMs, MM-LLMs, and Agentic AI into UAV systems, primarily targeting autonomy and multimodal perception. Examples include AirVista-II~\cite{airvistaII} for scene understanding, A4FN~\cite{A4FN} for automatic SLS generation and autonomous FN reconfiguration, multimodal LLM-enabled swarm coordination in dynamic environments~\cite{ping2025multimodallargelanguagemodelsenabled}, and surveys outlining Agentic UAV architectures~\cite{sapkota2025uavsmeetagenticai,Tian_2025}. In addition, frameworks such as ARMAIT~\cite{jiang2025agenticaiempoweredmultiuav} and FLUC~\cite{Framework_LLM_FALCON} leverage Agentic AI for trajectory optimization and mission control. Although these approaches demonstrate substantial progress toward autonomous aerial intelligence, they do not address the autonomous generation of structured outputs required for zero-touch network management in communications-oriented FNs.

Overall, state-of-the-art approaches predominantly rely on unimodal perception agents and remain disconnected from network control mechanisms. In contrast, FNs require a multimodal perception layer capable of fusing heterogeneous data sources—such as visual and audio information—to infer user distribution and traffic demand along the network operation. These limitations motivate the development of the proposed perception system.
\\

%% file: Chap_3/chap3.tex
\section{Multi-Agent Perception System}\label{chap:MAPS}
Enhanced situational awareness is essential to support autonomous decision-making in dynamic Flying Network (FN) environments. To enable this capability, we propose the Multi-Agent Perception System (MAPS), a modular and scalable perception system designed to provide FNs with autonomous environmental understanding. This section describes the MAPS architecture, including its core design principles, agent organization, and implementation.

\subsection{Concept}\label{Conceptual_Architecture}
\begin{figure}[]
  \begin{center}    \includegraphics[width=1.0\columnwidth]{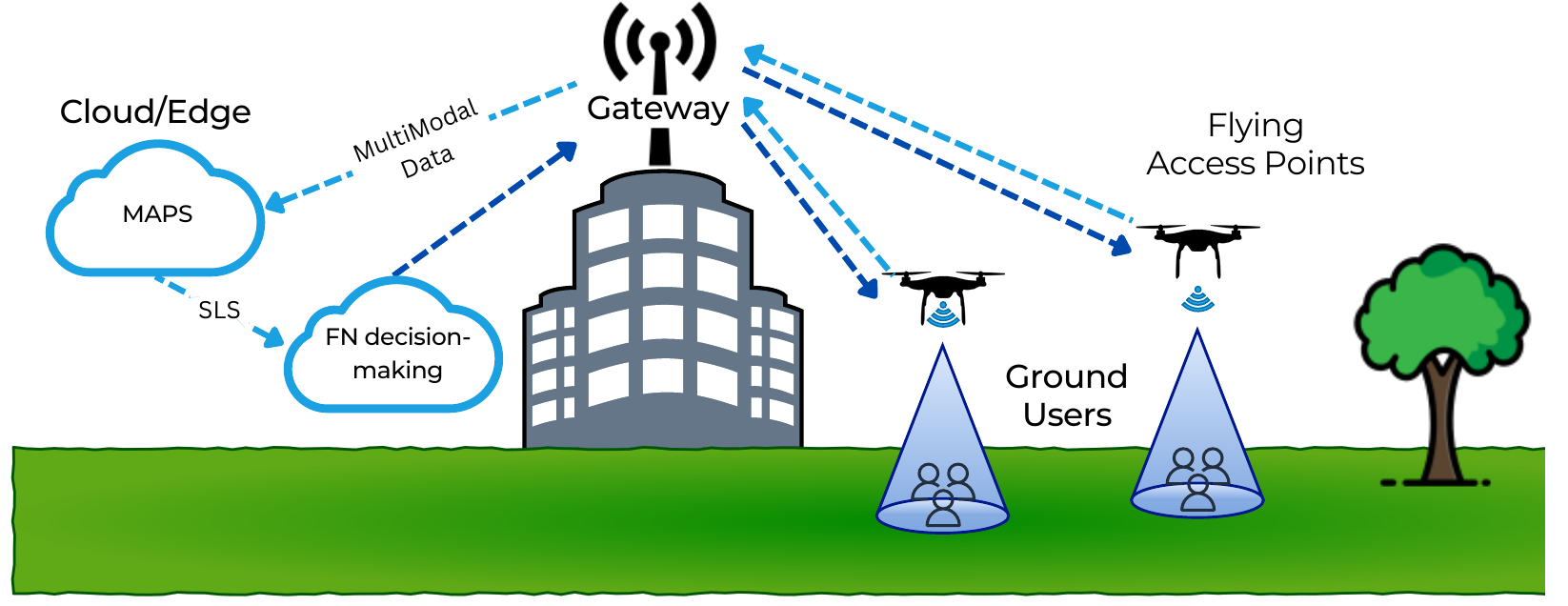}
    \caption{Reference FN architecture with MAPS deployed in cloud/edge computing. MAPS is responsible for providing the FN with the necessary data for FN decision-making. Adapted from \cite{1_ribeiro_supply_2024}.}
    \label{fig:FN_architecture}
  \end{center}
\end{figure}

MAPS combines Multi-Modal Large Language Models (MM-LLMs) and Agentic AI to generate SLSs, thereby providing the perception functionality required by the ETSI ZSM and ITU Autonomous Networks frameworks. It implements the perception layer envisioned in the A4FN architecture~\cite{A4FN} and builds upon the FN architecture proposed in~\textcite{1_ribeiro_supply_2024}, as depicted in Figure~\ref{fig:FN_architecture}. Through this integration, MAPS enables the network to interpret its operational environment and autonomously infer user needs.

MAPS provides the contextual information necessary for autonomous FN decision-making. Given the computational demands of MM-LLMs, MAPS is not designed for onboard deployment within UAVs but instead runs on edge or cloud computing infrastructure, enabling access to high-performance processing resources. Visual and audio data are transmitted from the UAVs to an external processing unit, where MAPS performs inference and data fusion before returning the SLS. This architecture balances inference latency with model capability, ensuring near real-time operation without compromising reasoning quality. 

\subsection{Design}

MAPS is designed as a modular, multi-agent solution that analyses multimodal inputs---including visual and transcribed audio data---to generate structured SLS outputs for use in autonomous FN operations. The system is guided by two primary objectives:
\begin{itemize}
    \item \textbf{Identification and labeling of users} in the operational scenario. This involves detecting and distinguishing individuals or entities, thereby estimating user count and their spatial distribution.
    \item \textbf{Estimation of traffic demand for each identified user.} MAPS leverages visual context and audio cues---such as a user’s proximity to an active fire or verbal indications of high data usage (e.g., image or video transmission)---to derive structured indicators of user traffic demand, supporting resource-aware FN decision-making.
\end{itemize}

\begin{figure}[]
  \begin{center}
    \includegraphics[width=0.6\columnwidth]{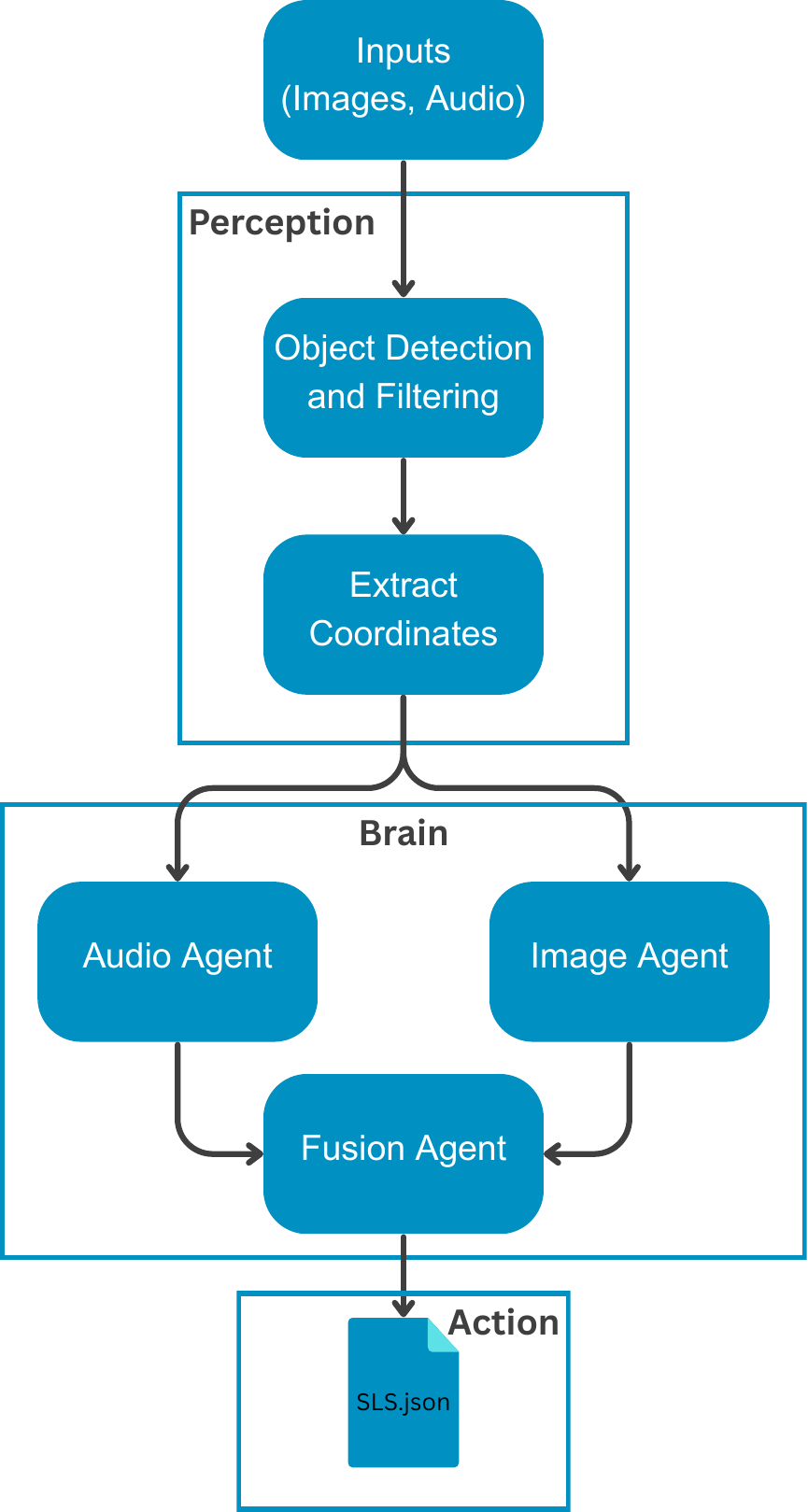}
    \caption{Overview of the MAPS architecture. The Perception layer pre-processes multimodal inputs; the Brain integrates three agents (Image, Audio, Fusion) for reasoning; and the Action layer generates structured Service Level Specifications (SLSs) for FN control.}
    
    \label{fig:MAPS architecture}
  \end{center}
\end{figure}

MAPS is organized into three main components, as depicted in Fig.~\ref{fig:MAPS architecture}: 1) \textit{Perception}, responsible for handling and pre-processing multimodal input data;  
2) \textit{Brain}, responsible for interpreting post-processed data and producing intermediate results through multiple agents; and  
3) \textit{Action}, represented by the generated SLS definition in a structured format (e.g., JSON file).

MAPS targets emergency scenarios, such as wildfires, and considers two primary data modalities: images and audio transcriptions collected from the operational environment. These modalities allow the analysis of user needs and the assessment of emergency intensity. Both data types follow a tailored processing pipeline designed to enhance relevant features and improve the accuracy and quality of LLM-based reasoning.

\subsection{Implementation}
Image processing begins with object detection to identify relevant entities in aerial imagery. To ensure accuracy and efficiency, MAPS employs YOLOv12~\cite{yolo12,tian2025yolov12}. Detections are filtered by object class and confidence thresholds: larger objects likely to contain users (e.g., cars, airplanes, trucks, buses) are retained if confidence exceeds 40\%, while smaller objects (e.g., persons, motorcycles) are retained if confidence exceeds 20\%. These thresholds were empirically selected, as lower values produced unreliable detections. For each valid detection, the center point of the bounding box is computed and stored as the relative coordinate of the detected user. The output of this stage is a structured file listing detected objects, corresponding $(x, y)$ coordinates, and an annotated image displaying labeled detections.

The \textit{Brain} component consists of multiple agents, each specialized in a specific task and input modality. It is structured as follows: the \textit{Image Agent} processes visual data, the \textit{Audio Agent} handles audio, and the \textit{Fusion Agent} combines the outputs of the first two agents into MAPS’ final output.

The \textit{Image Agent} analyses visual data, using object coordinates and bounding boxes to identify users and estimate their communication needs. It operates based on a structured prompt that frames it as a human operator interpreting UAV imagery, producing for each user a set of fields that describe spatial position, qualitative and quantitative traffic demand indicators, and a contextual justification in a standardized JSON format.

The \textit{Audio Agent} processes transcribed speech to extract intent and urgency. Using a similar structured prompt, it infers user needs that may not be visually evident, such as emergency calls or requests involving high-bandwidth data transfer. In real-world deployments, such transcripts may originate from first responders’ voice communications, captured using high-gain UAV-mounted microphones or receivers operating on the same channels as emergency teams. This capability enables inference of intent, urgency, and potential network load.

The \textit{Fusion Agent} merges the outputs of the Image and Audio Agents, resolving duplicates based on spatial proximity and synthesizing multimodal insights to produce a coherent and unified SLS. It applies cross-modal reasoning---for instance, increasing inferred traffic demand when both visual and audio cues indicate higher urgency---and ensures consistency and completeness in the final structured output used by FN decision-making entities.

Each agent employs an MM-LLM or single-modality LLM tailored to its function. In its current version, MAPS uses the Gemini 2.5 Flash model, accessed via an Application Programming Interface (API) and executed on Google Cloud infrastructure. This model outperforms its predecessor, Gemini 1.5 Pro, in all benchmarks reported by~\textcite{google_gemini_2_5_2025}. Leveraging cloud-based execution provides scalability and computational power while maintaining low-latency responses suitable for near real-time operation.

Given the multimodal nature of MAPS inputs, defining consistent and interpretable outputs across agents is essential. The structured output features of the Gemini model were used to enforce a JSON schema, ensuring reliable information extraction and formatting. The output of the \textit{Fusion Agent} represents the final MAPS output---a JSON-formatted SLS encapsulating the perceived environment in a structured representation. Each user entry includes the following fields:
\begin{itemize}
    \item \texttt{label}: An identifier that corresponds to the image tag.
    \item \texttt{x, y}: Relative spatial coordinates within the image frame.
    \item \texttt{throughput\_level}: A qualitative label indicating the throughput level, allowing an overseeing human to easily interpret the results (\textit{low}, \textit{medium}, \textit{high}).
    \item \texttt{context}: A natural language justification for the traffic demand, based on visual or audio cues.
    \item \texttt{traffic\_demand}: An estimate of traffic demand in Mbit/s.
\end{itemize}

The resulting SLS serves as an input for downstream algorithms performing UAV positioning, resource allocation, and network configuration, thereby enabling autonomous, zero-touch operation in FNs.

%% file: Chap_4/chap4.tex
\section{Multimodal Synthetic Data Generation}

To evaluate MAPS under realistic yet controlled conditions, a multimodal dataset was required, integrating synchronized audio and image data representative of emergency scenarios. While several single-modality datasets exist, such as the VisDrone dataset~\cite{droneviz} for aerial imagery, none specifically address the joint use of synchronized visual and audio modalities. Moreover, most available datasets are oriented toward general-purpose or surveillance contexts rather than mission-critical domains such as disaster response, which are the focus of this work. This lack of suitable multimodal data poses a significant limitation for the development and evaluation of perception systems such as MAPS.

To address this gap, we created a synthetic dataset of emergency scenarios designed to emulate realistic conditions encountered by FNs. The dataset was generated using Multi-Modal Large Language Models (MM-LLMs) to produce temporally aligned image and audio data. A set of thirty representative phrases commonly used by emergency personnel during field operations was developed, building on examples from the RescueSpeech dataset~\cite{RescueSpeech}. Each phrase generated was enriched with additional details describing user traffic demand and communication urgency, reflecting the contextual factors relevant to FN perception tasks.

\begin{figure}[!ht]
    \centering
    \subfloat{{\includegraphics[width=0.95\columnwidth]{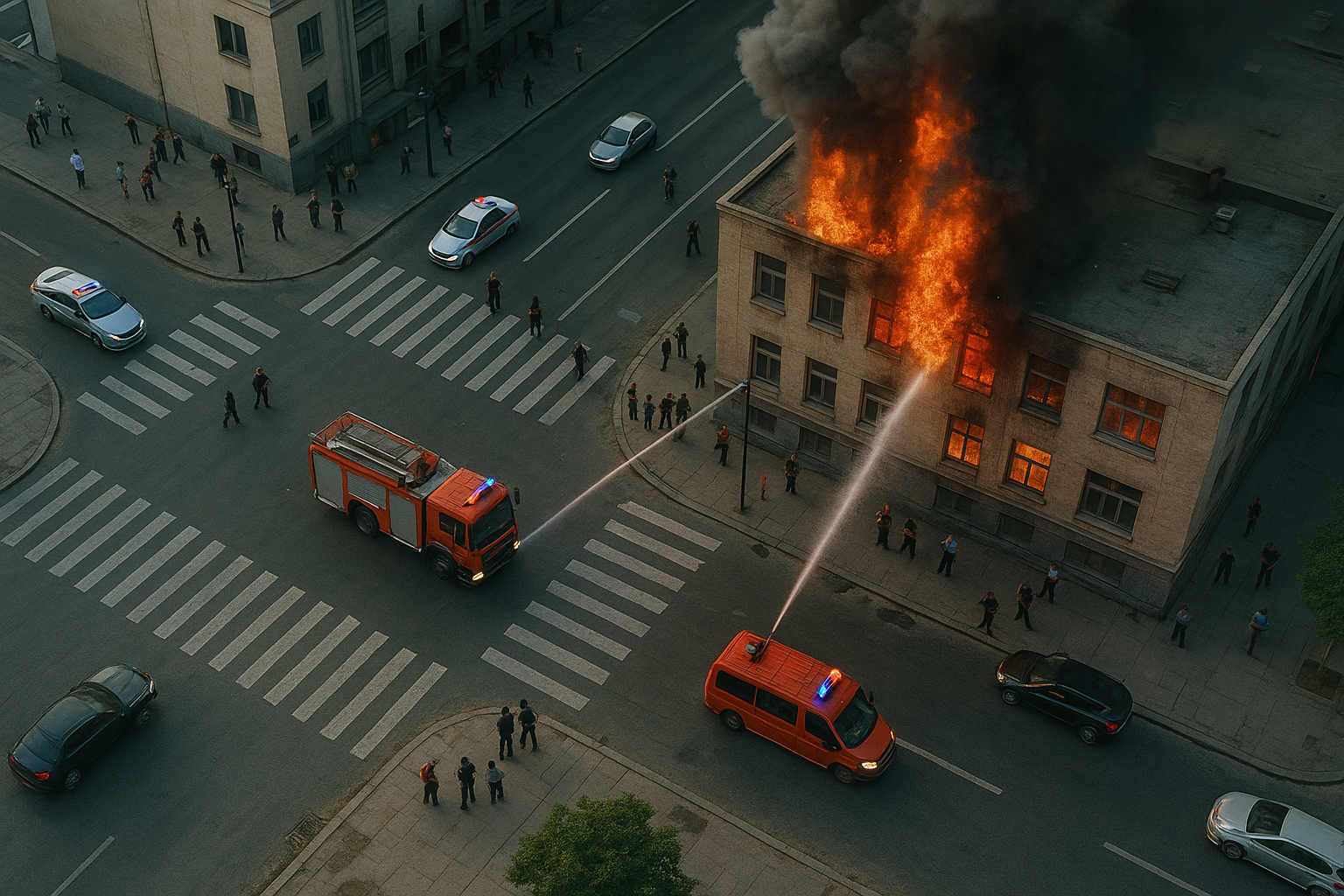}}}
    \vspace{0.03\columnwidth}
    \subfloat{{\includegraphics[width=0.95\columnwidth]{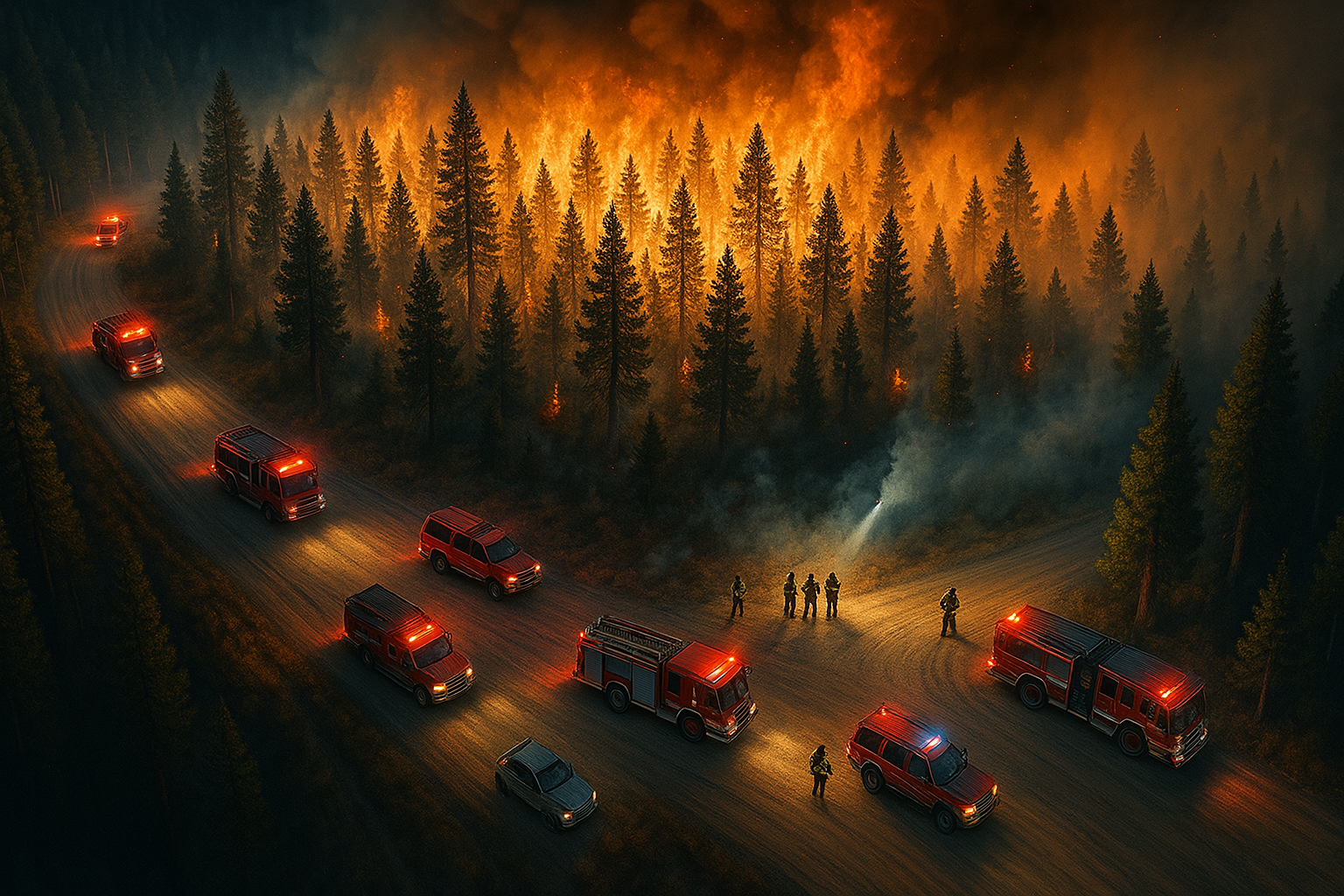}}}
    \caption{Examples of synthetically generated emergency scenarios from the dataset available in~\cite{Cardona_MAPS-Dataset_2025}, illustrating disaster management scenes. The dataset combines visual and audio modalities for MAPS evaluation.}
  \label{fig:Examples_used_images}
\end{figure}

A dedicated prompt was then engineered to guide the generation of images depicting diverse emergency situations---including wildfires, search and rescue operations, and disaster management scenes---as illustrated in Fig.~\ref{fig:Examples_used_images}. Prompt design followed best practices in LLM prompt engineering described by~\textcite{boonstra2024prompt}, ensuring detailed scene composition and semantic alignment between visual and audio elements. Image generation was performed using the DALL·E~3 model~\cite{openai_dalle3_image_2025}.

Following image generation, each sample was manually annotated to create the dataset’s ground truth. Since this paper focuses on FN perception in emergency contexts, the ground truth includes bounding boxes around visible users and objects likely to contain users, enabling precise spatial evaluation of user detection. To incorporate the audio modality, corresponding dialogues were synthesized by referencing the previously defined set of emergency phrases and analyzing each scenario to ensure semantic consistency between speech and visual context. A complementary prompt was designed to facilitate this multimodal synthesis.

The resulting dataset comprises 20 distinct scenarios, each providing synchronized visual and audio content with varying levels of complexity and user density. This diversity ensures robust testing and performance evaluation of MAPS across a
range of operational conditions. The dataset is publicly available in~\cite{Cardona_MAPS-Dataset_2025}, supporting reproducibility and further research on multimodal perception for UAV-assisted communication systems. Beyond MAPS evaluation, this dataset aims to catalyze research on multimodal UAV perception, offering a reproducible baseline for future works.

\section{System Evaluation}

MAPS was evaluated in terms of detection accuracy and execution time to assess its suitability for near real-time operation in autonomous FNs.

\subsection{MAPS Accuracy}
A key performance indicator for MAPS is its accuracy in detecting users and interpreting the surrounding environment. This subsection provides a quantitative assessment of this indicator across several operational scenarios.

The accuracy is defined as:

\begin{equation}
\label{eq:accuracy}
\text{Accuracy} = 
\frac{N_{\text{detected}}}{N_{\text{ground\_truth}}} \times 100~[\%]
\end{equation}

\noindent
where $N_{detected}$ and $N_{ground\_truth}$ represent the number of users detected by MAPS and the number of users manually annotated, respectively.

\begin{figure}[]
  \begin{center}
    \includegraphics[width=1\columnwidth]{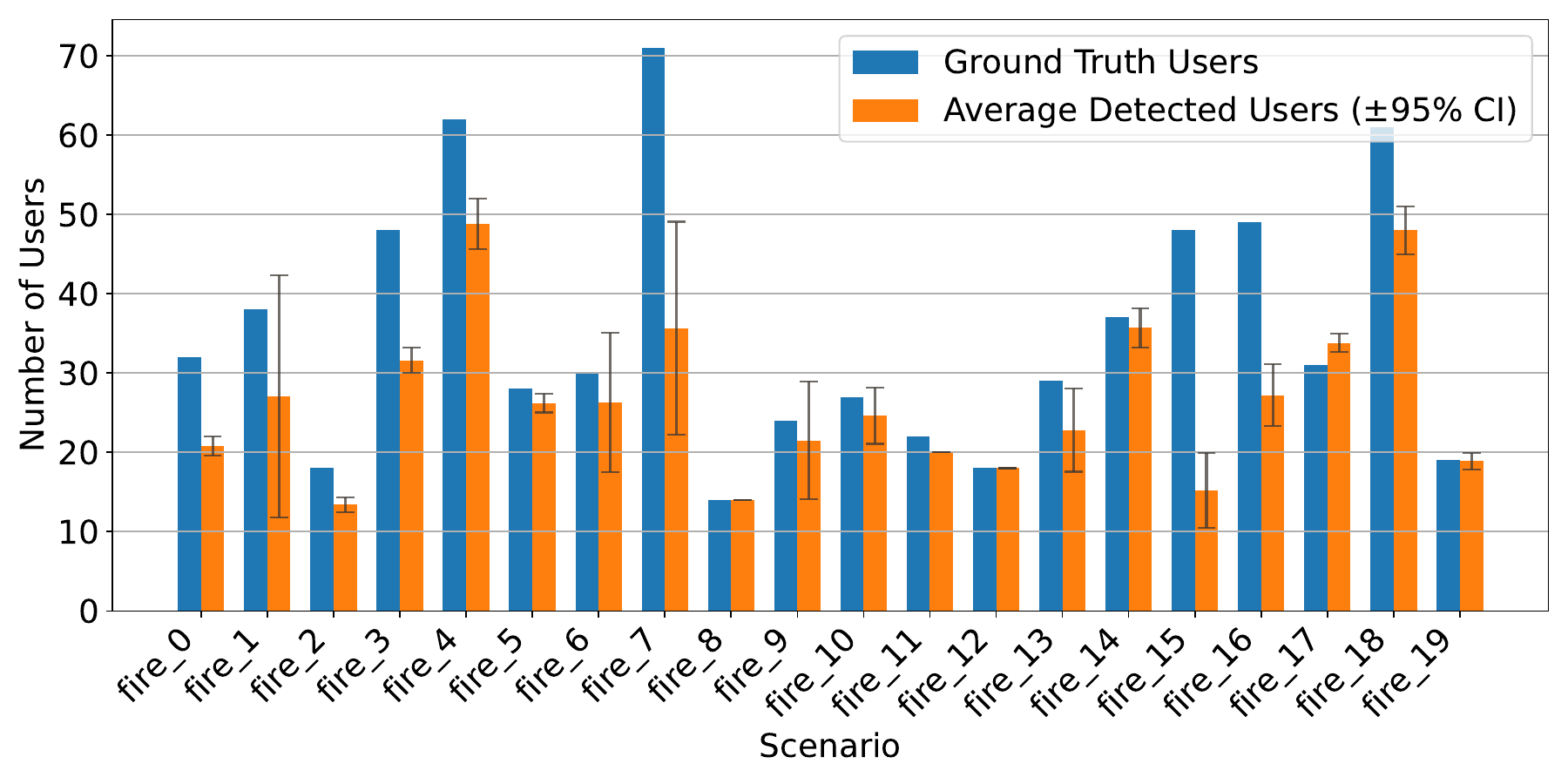}
    \caption{Comparison between the number of users detected by MAPS and the ground truth across all scenarios, considering 10 runs per scenario. The ground truth includes both visible individuals and objects typically associated with user presence (e.g., vehicles).}
    \label{fig:Number of users detetcted}
  \end{center}
\end{figure}

Figure~\ref{fig:Number of users detetcted} compares the average number of users detected by MAPS with the ground truth across 20 scenarios. The results demonstrate strong alignment between detections and actual values, particularly in mid-range scenarios (15--30 users), where both accuracy and consistency remain high. 

In 70\% (14 out of 20) of the evaluated scenarios, MAPS achieved user detection accuracy above 70\%, with three scenarios reaching 100\%. A smaller subset of scenarios---\textit{fire\_1}, \textit{fire\_7}, and \textit{fire\_15}---showed accuracy below 60\%, primarily due to poorly defined or visually ambiguous objects in the synthetic images, especially in densely populated scenes. This limitation is expected to be mitigated when applying MAPS to real-world imagery, where object quality and spatial features are more consistent.

\begin{figure}[]
  \begin{center}
    \includegraphics[width=1\columnwidth]{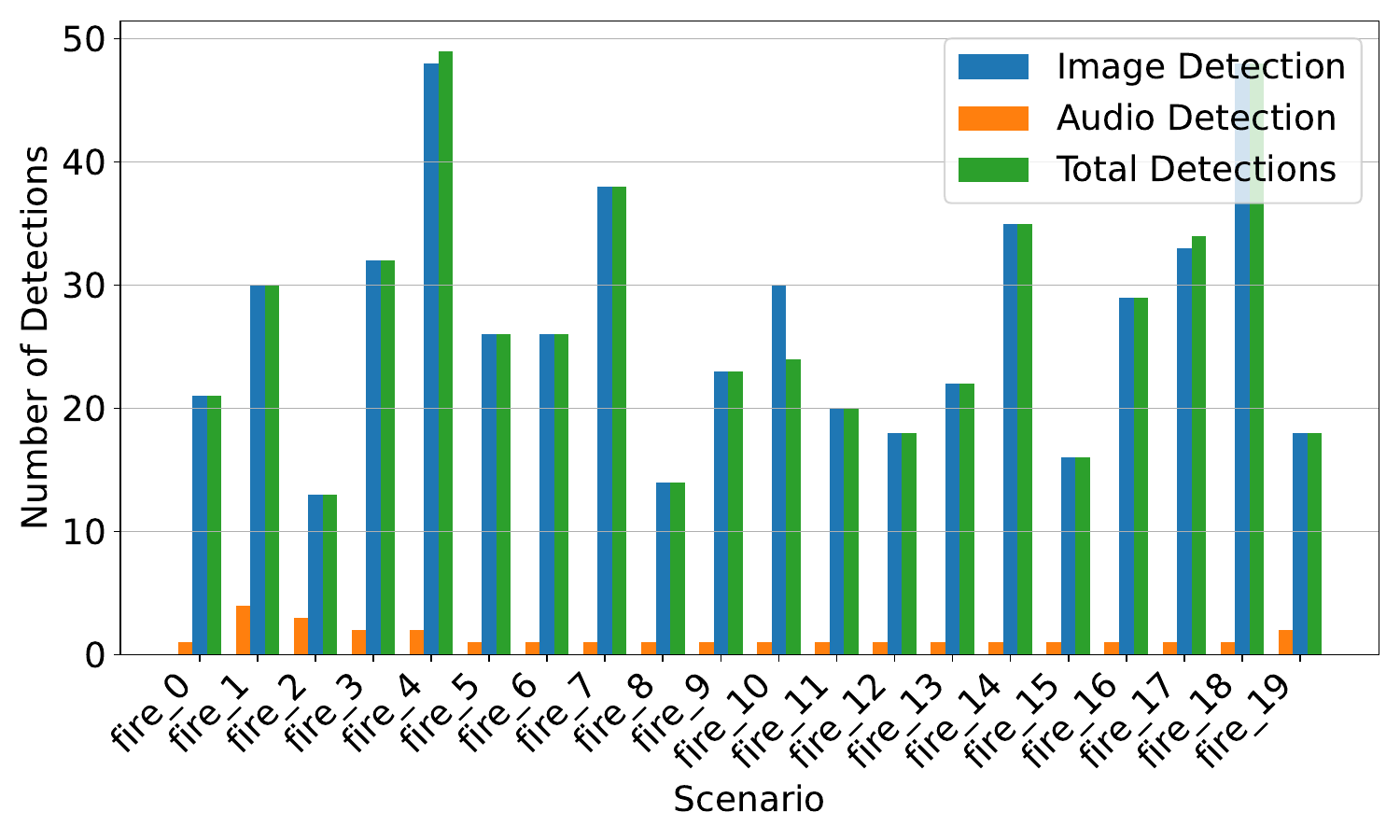}
    \caption{Number of users detected by each MAPS agent (Image, Audio, and Fusion) in one representative run. The results illustrate the contribution of each modality to the overall detection process and were consistent across multiple runs conducted to ensure statistical reliability.}
    \label{fig:number of users detected per component}
  \end{center}
\end{figure}

Figure~\ref{fig:number of users detected per component} presents the number of user detections produced by each MAPS component: the \emph{Image Agent}, \emph{Audio Agent}, and \emph{Fusion Agent}. The \emph{Image Agent} is the dominant contributor, accounting for the vast majority of detections across all scenarios. The \emph{Audio Agent} typically detects one or two users, suggesting a minor yet complementary role in the current system configuration. In most cases, total detections closely match the \emph{Image Agent} output, with slight increases in some scenarios indicating that the fusion process successfully incorporates supplementary detections from the audio modality. In rare cases, such as \textit{fire\_10}, the \emph{Image Agent’s} large language model produced an overestimation of total users, which the \emph{Fusion Agent} subsequently corrected by resolving inconsistencies across modalities.

These results confirm that MAPS primarily relies on vision-driven perception while maintaining multimodal adaptability. The integration of audio cues enhances detection robustness in specific conditions, validating the feasibility of multimodal perception as a foundation for zero-touch operation in FNs.

\subsection{MAPS Execution Time}
\begin{figure}[]
  \begin{center}
    \includegraphics[width=0.95\columnwidth]{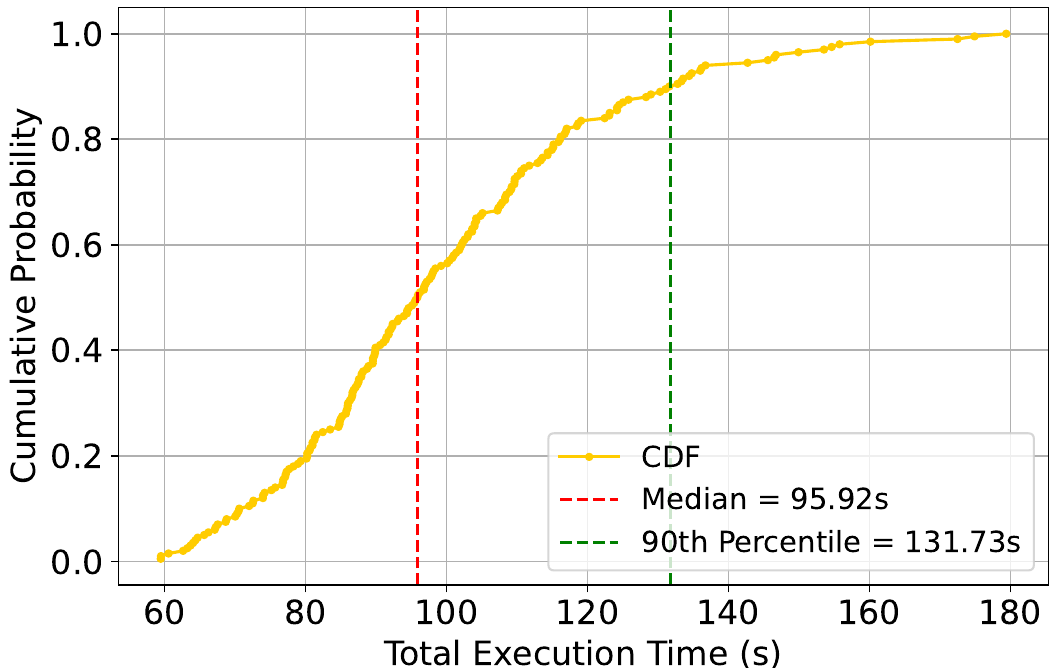}
    \caption{Cumulative distribution of MAPS execution time across 20 synthetic emergency scenarios. The results show that 90\% of executions are completed within 132~s, demonstrating the system’s suitability for near real-time operation.}
    \label{fig:CDF_EXEC}
  \end{center}
\end{figure}

Execution time is a critical performance metric for near real-time or time-sensitive FN operations. Although MAPS employs complex multimodal reasoning through large language models, it must remain within acceptable latency bounds to be deployable in practical environments. To evaluate responsiveness, MAPS was executed across the 20 synthetic emergency scenarios. As shown in Fig.~\ref{fig:CDF_EXEC}, 90\% of executions completed in under 132~s, with most executions ranging between \SI{80}{\second} and \SI{120}{\second}, as indicated by the steep cumulative slope in that interval.

Although the overall latency is compatible with near real-time operation, a more detailed breakdown is required to identify bottlenecks and guide optimization. A profiling-based approach, typically used in heterogeneous system analysis, was employed to measure the time contribution of each processing stage. The \texttt{cProfile} module in \textit{Python} was used to collect detailed runtime statistics on function calls, including execution counts and cumulative processing times.

\begin{figure}[]
  \begin{center}
    \includegraphics[width=1\columnwidth]{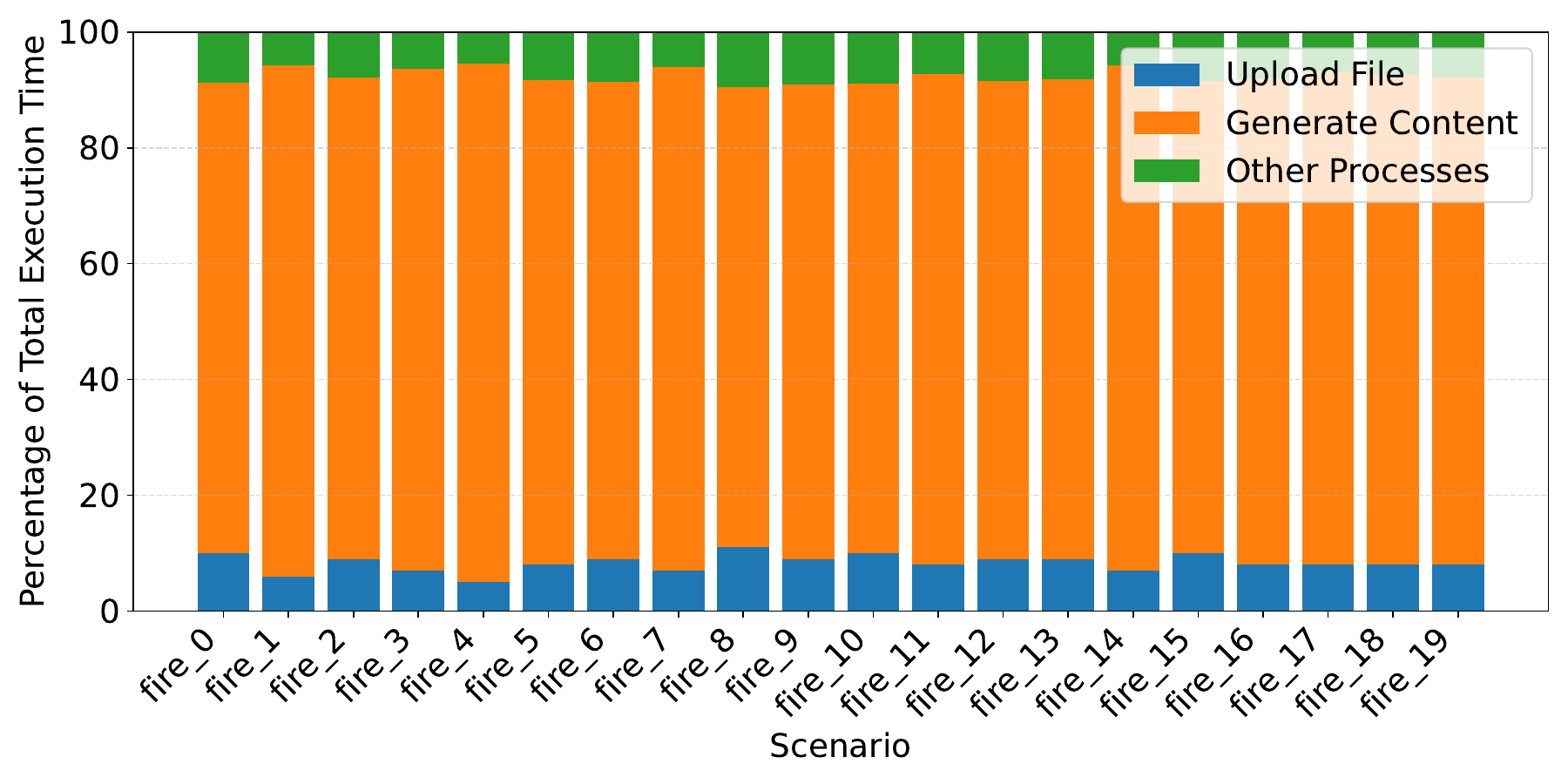}
    \caption{Percentage of total execution time spent on the most time-intensive MAPS functions across the 20 evaluated scenarios. The results show that the \textit{Generate Content} component dominates the execution pipeline, accounting for approximately 80\% of total latency.}
    \label{fig:Programs_distribution}
  \end{center}
\end{figure}

Figure~\ref{fig:Programs_distribution} shows the distribution of execution time across key MAPS components. The \textit{Generate Content} function---responsible for sending requests to Gemini’s API and processing structured responses---is the dominant contributor, accounting for approximately 80\% of total execution time. This function is executed three times per MAPS cycle, once for each agent (\emph{Image}, \emph{Audio}, and \emph{Fusion}). The consistent dominance of this component indicates that API interaction latency constitutes the primary performance constraint.

In contrast, the \textit{Other Processes} component---which includes image preprocessing, coordinate extraction, and data structuring---along with the \textit{Upload Files} component, contribute minimally to the total latency. These results indicate that MAPS’s low-level perception and preprocessing stages are already well optimized, and further performance improvements should focus on reducing language model interaction time and optimizing API-level efficiency.

%% file: Chap_last/Conclusion.tex
\section{Discussion \label{sec:Discussion}}

The evaluation results confirm that MAPS achieves high detection accuracy and rapid inference performance, representing a significant step toward fully autonomous FNs. The system demonstrates that combining perception, reasoning, and structured output generation within a multi-agent architecture enables effective environmental awareness for zero-touch operation.

Most scenarios achieved detection rates above 70\% for targets of interest, validating the reliability of the underlying perception pipeline. Performance degradation was observed only in a few cases, primarily due to the low visual fidelity of some synthetically generated images. These artifacts affected the object detector’s ability to correctly identify users or vehicles. In practical deployments, where imagery can be captured under more stable lighting and resolution conditions, such degradation is expected to be less prominent. The results also confirm that MAPS remains predominantly vision-driven, as the \emph{Image Agent} provides the majority of detections, while the \emph{Audio Agent} contributes incremental refinements that enhance robustness in specific contexts.

Integrating the audio modality improved both user detection and traffic demand estimation; however, the impact was limited by the small number of participants represented in the synthetic conversations. Expanding the dataset to include additional audio samples and more complex dialogues would likely yield further gains in multimodal reasoning accuracy and the precision of generated Service Level Specifications (SLSs). Moreover, retraining or fine-tuning the object detection component on domain-specific emergency imagery could strengthen visual reliability and reduce the observed sensitivity to synthetic data quality.

The execution time analysis demonstrated that MAPS operates within latency bounds suitable for near real-time FN deployment. Although FN environments are dynamic, they typically evolve over time scales that allow for periodic perception updates. Thus, MAPS can be executed at regular intervals to maintain situational awareness and support adaptive reconfiguration of UAV positions and network resources. This periodic execution model balances system responsiveness with computational efficiency.

The profiling results highlighted that the \textit{Generate Content} function, responsible for communication with the Gemini API, dominates overall latency, accounting for approximately 80\% of total execution time. Optimization efforts should therefore prioritize reducing Large Language Model (LLM) interaction delays. Potential strategies include prompt simplification, parallel execution of agent tasks, and batching of contextual data to minimize redundant API calls. In addition to software-level improvements, latency could also be mitigated by adopting alternative cloud-based LLMs that offer lower response times or by deploying lighter-weight models at the network edge, where processing occurs closer to data sources. Both approaches could significantly enhance MAPS’s responsiveness and further improve its suitability for time-critical FN operations.

In a nutshell, the results indicate that MAPS effectively bridges the gap between multimodal perception and autonomous decision-making, while the identified limitations provide clear directions for future work.

\section{Conclusions and Future Work\label{chap:Conclusions}}

This paper presented MAPS, an AI-based multi-agent perception system designed to enable autonomous environmental awareness in FNs. MAPS addresses a critical limitation of existing FN approaches—the lack of autonomous perception—by integrating multimodal data analysis into a modular, multi-agent architecture built upon Multi-Modal Large Language Models. The system autonomously interprets aerial imagery and audio inputs to generate structured Service Level Specifications (SLSs) that support zero-touch decision-making in FN management.

Experimental evaluation using a synthetic emergency dataset demonstrated that MAPS can complete perception and SLS generation in under 130~seconds in 90\% of cases, while achieving user detection accuracy above 70\% in most of the evaluated scenarios. These results confirm that MAPS operates within the latency and accuracy bounds required for near real-time FN deployments. The analysis also revealed that the system’s performance is primarily vision-driven, with audio inputs providing complementary context that enhances perception robustness in selected scenarios. However, performance may be constrained by the visual quality of imagery and by LLM inference latency when relying on external cloud APIs.

Future work will focus on four directions: 1) expanding the multimodal dataset with real-world captures; 2) enriching SLS generation through additional telemetry and network data; 3) deploying MM-LLMs at the network edge to reduce inference latency; 4) validating MAPS in real-world testbed.

\section*{Acknowledgements}
This work is part of the FALCON project (10.54499/2023.15645.PEX), funded by the Portuguese Foundation for Science and Technology (FCT).